# Techno-economic analysis of self-sustainable thermophotovoltaic systems for grid-scale energy generation

Jihun Lim[1,2,3†] and Sungwon Lee[1],

[1]Department of Semiconductor Science and Technology, Jeonbuk National University, Jeonju, 54896 Republic of Korea

[2]School of Semiconductor and Chemical Engineering, Jeonbuk National University, Jeonju, 54896 Republic of Korea

[2]Research Institute for Materials and Energy Sciences, Jeonbuk National University, Jeonju, 54896 Republic of Korea

[†]Correspondence email: jihunlim@jbnu.ac.kr

## Abstract

To facilitate the widespread adoption of renewable energy, dispatchable, zero-emission power sources are essential for grid stability. This work performs a comprehensive techno-economic analysis of a self-sustainable thermophotovoltaic (TPV) system, an architecture that integrates solar charging to function as a standalone power generation asset. Using theory-based models for conventional air-bridge InGaAs and Si diode cells, our analysis reveals that while the system is not currently competitive from a pure levelized of storage cost (LCOS) perspective due to the high capital expenditure for thermal battery materials, its primary value lies in its competitive levelized cost of electricity (LCOE), which is comparable to that of conventional dispatchable generators such as gas turbines. Furthermore, we show that a full Si-based TPV system, utilizing a 50-μm-thick air-bridge cell for enhanced photon utilization, can also achieve an LCOE that is competitive with such conventional power sources at scales exceeding the gigawatt-hour level, despite its lower conversion efficiency relative to its InGaAs counterpart. This highlights a practical engineering pathway for leveraging the immense manufacturing scalability of Si, offering a lower-risk route to deployment compared to III-V materials. Ultimately, this work establishes the self-sustainable TPV architecture as a compelling pathway toward providing grid-scale, on-demand, zero-emission power.

**Introduction**

Advanced technologies such as electric vehicles, artificial intelligence, and robotics are driving a significant increase in global electricity demand [1, 2]. Traditional power generators such as turbine-based facilities and natural gas combustion systems still account for nearly half of worldwide electricity production [3]. This growing demand places increasing strain on existing energy grids, many of which were not originally designed to accommodate high-density, distributed, or renewable energy inputs. In response, carbon neutrality has emerged as a global priority, prompting accelerated development of renewable and sustainable energy technologies aimed at reducing greenhouse gas emissions and modernizing power infrastructure [3, 4]. Solar and wind systems are among the most widely adopted eco-friendly electricity generation technologies [5], currently producing more than 4,000 TWh per year globally [6]. In particular, solar systems have been broadly implemented not only in industry, but also in residential and commercial sectors. However, their performance is constrained by climate conditions, installation geometry, and time-dependent availability, limiting their contribution to less than 5 % of total electricity generation [5, 7].

In contrast to PV systems, thermophotovoltaic (TPV) systems can serve as independent power generation resources by directly converting thermal radiation into electricity, regardless of climate changes or time of day. Recent advancements in back-surface-reflector (BSR)-embedded TPV cells have demonstrated heat-to-power conversion efficiencies exceeding 40% [8-10], rivaling those of existing power generators such as coal steam turbine (30-40%), nuclear power plants (30-35%), wind turbines (35-50%), solar PV modules (15-25%), and diesel engines (35-40%) [11-13]. Single-junction TPV cells based on $In_{0.53}Ga_{0.47}As$ (InGaAs) positioned at close distances – typically from a few millimeters down to hundreds of micrometers – from thermal emitters, have achieved relatively high output power densities ($P_{out}$) greater than 3 W/cm$^2$ [14] that one hundred times higher than conventional Si solar PV cells (< 0.02 W/cm$^2$) [15, 16]. For instance, Tervo *et al* demonstrated Au-BSR InGaAs TPV arrays on a 3-inch Si wafer, highlighting the scalability potential of conventional III-V TPV module [14]. However, III-V TPV technologies yet rely on smaller wafer sizes than Si and involve costly, time-consuming processes such as lattice-matched epitaxial growth and the sacrificial use of InP substrates, all of which restrict their potential for large-scale integration and mass production.

Si, which offers a fundamentally more manufacturable and scalable platform [17, 18], requires further study to improve efficiency and overcome key limitations. To maximize the conversion of solar energy, extensive research has been devoted to advanced system configurations. For instance, sophisticated approaches such as full-spectrum splitting and residual-spectrum reshaping have been explored to utilize the entire solar irradiance efficiently in novel cascade photovoltaic systems [19-21]. Concurrently, significant efforts have focused on optimizing the core components of concentrated solar thermophotovoltaic (CSTPV) systems [22-24]. It is crucial to note, however, that advanced solar cell structures are not always directly transferable to TPVs. Unlike solar cells managing the broad solar spectrum, TPV systems prioritize optimizing a controllable heat source. Consequently, for a TPV cell with a back reflector, the critical task is to enhance photon recycling via high reflectance and low parasitic absorption. These strategies often rely on complex optical arrangements or expensive, non-silicon-based high-bandgap cells, which can pose challenges for economic viability and large-scale deployment. A critical research gap remains in developing a TPV system that combines high performance with the proven scalability and low-cost advantages of silicon-based manufacturing. In 2021, Lee *et al.* reported a 4 mm × 4 mm air-bridge 500-μm thick Si TPV cell featuring a lateral electrode structure [25], which exhibited performance limitations due to significant surface recombination and a high series resistance ($R_s$) above 200 mΩ·cm². As a result, the cell only demonstrated a conversion efficiency of 19 % and output power density of 0.06 W/cm². However, a clear pathway for demonstrating such high-performance Si TPV cells that are both commercially viable and scalable for TPV applications has yet to be established.

In this work, we perform a comprehensive techno-economic analysis to evaluate the viability of a self-sustainable TPV system. The analysis considers two distinct material platforms: established high-performance InGaAs cells and emerging low-cost Si cells. The performance characteristics of these cells are established through physics-based methodologies, including technology computer-aided design (TCAD) and diode modeling, validated with experimental principles from transfer matrix method (TMM) and Fourier-transform infrared (FTIR) spectroscopy. To create a meaningful comparison, we establish the key design principles required for a Si TPV cell to close the performance gap with its InGaAs counterpart, specifically assuming a thinned wafer architecture and a low metal contact resistivity ($R_c$). Based on these optimized cell designs, we evaluate the system-level levelized cost of energy (LCOE) and benchmark it against conventional power generation technologies, including PV, CSP, and gas turbines. This study not only assesses the current economic standing of these TPV systems but also identifies the most critical technological hurdles and presents a practical pathway for the future development of large-scale, zero-emission, dispatchable TPV power generation.

**Enhanced Optical and Electrical Performance of a Thin Si Thermophotovoltaic Cell**

Figure 1(a) illustrates the configuration of the comprehensive TPV system analyzed in this work. The system consists of a TPV cell module, charging stages such as solar PV and concentrated solar power (CSP), and a Si thermal battery. In the CSP-TPV system, concentrated solar energy is directed onto the outer surface of the SiC vessel, enabling efficient heat transfer to the internal silicon thermal battery via conduction and radiation. Electricity from the PV system powers an electric heater embedded within the cylinder, providing direct heating to the silicon. The detailed techno-economic analysis of this integrated system will be presented in a later section. The core power-generating component within the TPV module of this system is either an InGaAs or a Si TPV cell. While InGaAs TPV cells have demonstrated high conversion efficiencies approaching 40% with power densities more than 1 W/cm$^2$ [8, 14], Si TPV cells have failed to achieve comparable performance. Therefore, a technological breakthrough in Si TPV cell design is required to overcome these limitations and establish it as a viable low-cost alternative for TPV applications.

Figure 1(b) shows the unit structure of a Si TPV cell featuring vertically aligned contact gridlines and an air-bridge reflector. The Si absorber thickness ($t_{Si}$) and carrier concentration critically influence photon dynamics in the out-of-band (OOB) region, where the air-bridge structure enhances photon reflection at the cell bottom [10, 26, 27]. The conversion efficiency ($\eta_{TPV}$) is defined as [26]:

$$\eta_{TPV} = \frac{P_{out}}{P_{abs}} = \frac{P_{out}}{P_{out}+P_{diss}} = SE \cdot IQE \cdot V_F \cdot FF,$$

where *SE* is spectral efficiency, *IQE* is internal quantum efficiency, $V_F$ is the voltage factor = $V_{oc}/E_g$, and *FF* is the fill factor. The $P_{diss}$ is the dissipated power, which includes parasitic photon absorption. This parasitic absorption is reduced by the air bridge architecture, which enhances OOB reflectance ($R_{OOB}$) and *SE* [27, 28]. With an absorber carrier concentration of $10^{16}$ /cm$^3$, the top and bottom panels in Fig. 1(c) present the simulated absorption spectra and $R_{OOB}$ for Si membranes with different $t_{Si}$ (10 μm, 50 μm, 200 μm, and 500 μm) freestanding on a substrate composed of a 600-μm air gap, a 1000-nm Au layer, and a Si wafer. The simulations were performed using the transfer matrix method, incorporating free carrier absorption (*FCA*) via an experiment-based Drude model ($\alpha_{FCA}=CN\lambda^\gamma$ where *C*: $10^{-10}$ and $\lambda$: 2) [10, 26, 29]. Due to FCA, the $R_{OOB}$ continuously drops with increasing wafer thickness, falling below 92 % at $t_{Si}$ = 500 μm. This thickness-dependent trend was experimentally investigated using FTIR measurements, as detailed in Supplementary Information A.

Figure 1(d) represents that reducing the absorber thickness enhances $R_{OOB}$ as a function of blackbody temperatures ($T_{BB}$). The inclusion of *FCA* in Si leads to significant absorption at higher energies, resulting in that $R_{OOB}$ can be below 95% for a relatively thick Si. For $t_{Si}$ < 150 μm, the $R_{OOB}$ is enhanced more than 98 % that is competitive to InGaAs-based TPV cells [8, 10, 27, 30]. Figure 1(e) shows the corresponding *SE* versus $T_{BB}$, where *SE* depends on the in-band absorption and $R_{OOB}$. Air-bridge InGaAs TPV cells typically have reported *SE* above 70% [8, 10, 27, 30]. As highlighted by the dashed line, a Si TPV cell with $t_{Si}$ near 50 μm can achieve comparable *SE* at $T_{BB}$ > 1650 °C (unity view factor) to that of InGaAs TPV cells.

Figure 2(a) presents a comparison of representative current density-voltage (*J-V*) characteristics for Si cells ($t_{Si}$ = 200 μm), simulated using both a TCAD suite (dashed line) (the air-bridge area fill factor of 87.5 %) and a previously developed TPV diode model (solid line)

[26] under illumination. The open-circuit voltage ($V_{oc}$) and short-circuit current density ($J_{sc}$) were calibrated to match those of a conventional Si solar cell ($V_{oc}$ = 0.67 V, $J_{sc}$ = 0.036 A/cm$^2$, $FF$ = 82.1%, $P_{out}$ = 0.020 W/cm$^2$) [26, 31]. The shaded region indicates the performance variation for Si TPV cells. This variation arises from differences in non-radiative and radiative recombination due to process corners, benchmarked against various Si solar cell performance data from the literature [25, 26, 31-35]. Details of the TCAD and diode models are provided in Supplementary Information B and C. Owing to the computational demands of TCAD simulations, the TPV diode model was used for extensive $J$-$V$ calculations across various $T_{BB}$ and heatsink designs. Decreasing $t_{Si}$ reduces the photocurrent and, consequently, the output power ($P_{out} = J_{sc} \cdot V_{oc} \cdot FF$). This trend is in contrast to the enhanced $R_{OOB}$ and $SE$ observed in Fig. 1, underscoring the critical importance of optimizing $t_{Si}$ for maximum conversion efficiency. The TCAD simulation assumes a metal contact resistivity ($R_c$) to p-type Si of 0.98 mΩ·cm$^2$, which falls within the typical range reported for Si cells [31, 36-40]. This results in the $R_s$ set to 80.6 mΩ·cm$^2$, extracted from the measurement using a Si cell (Figure S3(d) in Supplementary Information). The interface between p-type Si and metal is susceptible to Fermi-level pinning and inefficient charge tunneling, resulting in higher $R_s$ in Si diodes [41, 42]. By contrast, InGaAs TPV cells typically exhibit $R_s$ values of 10 mΩ·cm$^2$ or less [9, 10, 14].

Figure 2(b) shows the dependence of $V_F \cdot FF$ product, representing carrier management in Si TPV cells, as a function of $R_s$ and $T_{BB}$ with $t_{Si}$ fixed at 50 μm. For these simulations, the Si absorber thickness was reduced to 50 μm for optimizing $SE$ as previously shown in Fig. 1(d). The high $J_{sc}$ at elevated temperatures drives the diode into the series-resistance-limited regime, rendering $FF$ and $P_{out}$ highly dependent on $R_s$. $P_{diss}$, which includes heat from non-radiative recombination, parasitic absorption (via FCA and optical interference), and Joule heating. These directly affect the junction temperature ($T_J$) and the self-heating effect is accounted for the *J-V* simulations [26]. Figure S4 (Supplementary Information B) provides the simulated $T_J$ with various simulation conditions. The range of $V_F \cdot FF$ for InGaAs TPV cells is 0.44 – 0.53 [8, 10, 27, 30] with different view factor and emission temperatures. The $V_F \cdot FF$ decreases as $T_{BB}$ and $R_s$ increase due to Joule heating, which causes a reduction in *FF*. Using the air-bridge InGaAs cell model (Fig. S5(a) in Supplementary Information C), the $V_F \cdot FF$ is expected to be in the range of 0.28 – 0.55. Notably, Si TPV cells with $R_s$ below 30 mΩ·cm² exhibits the $V_F \cdot FF$ product that resides firmly within the minimum and maximum boundaries (marked by the dashed lines) of the InGaAs cell model across the studied emitter temperatures.

Figure 2(c) presents the conversion efficiency of the Si TPV cell as a function of $T_{BB}$ and $R_s$, where the *IQE* is assumed to be 98 % [43, 44]. The dashed lines represent efficiency contours of 25 %, 30 %, and 35 %. With a practical $R_s$ of 80 mΩ·cm², the Si TPV cell achieves a maximum efficiency of 29 % at $T_{BB}$ = 1300 °C. Reducing $R_s$ to 30 mΩ·cm² and 10 mΩ·cm² increases the maximum efficiency to 34 % (at $T_{BB}$ = 1400 °C) and 38 % (at $T_{BB}$ = 1700 °C), respectively. The ~39 % maximum efficiency of InGaAs cells (Fig. S5(c) in Supplementary Information) highlights that achieving a low $R_c$ at the p-type Si/metal interface remains the key challenge for high-performance Si TPV cells.

Figure 3(a) shows the figure of merit (conversion efficiency) for InGaAs-based TPV cells (symbols) from the literature, where the dashed lines indicate the efficiency levels from 25% to 50%. The two trajectory lines predict Si TPV cell performance as a function of $R_s$ (80 mΩ·cm² to 5 mΩ·cm², with $t_{Si}$ = 50 μm) and $t_{Si}$ (10 μm to 500 μm, with $R_s$ = 5 mΩ·cm²), presenting optimized conversion efficiency across $T_{BB}$ from 1000 °C to 1800 °C. InGaAs TPV cells fall between 35% and 40%, whereas Si TPV cells can reach this range when $R_s$ is below 30 mΩ·cm². Figure 3(b) presents the simulated output power density of the Si TPV cell ($t_{Si}$ =

50 μm) as a function of $R_s$ and $T_{BB}$. For $R_s$ above 80 mΩ·cm², the maximum power density remains below 2 W/cm²; as $R_s$ approaches 30 mΩ·cm² or lower, the power density can exceed 4 W/cm². Figure 3(c) plots power density versus conversion efficiency, comparing experimental results for InGaAs TPV cells [8, 14, 27, 45] and simulation results for the Si TPV cell. The solid line shows the trend for the Si TPV cell as $R_s$ varies from 80 mΩ·cm² to 5 mΩ·cm²; the dashed line indicates the improvement associated with $V_F$ [26]. Power densities were simulated at optimized conversion efficiency values for $T_{BB}$ in the range of 1000°C and 2000°C.

**Techno-economic analysis for the thermal battery-based TPV system**

Figure 4(a) illustrates the conceptual schematic of a thermal battery, where Si serves as the phase change material contained within a SiC vessel. Upon heating and melting, Si stores both sensible and latent thermal energy, and the SiC surface emits radiative heat to the surrounding TPV module. The total stored thermal energy depends on the Si volume, determining both the thermal battery size and the corresponding area of InGaAs or Si-based TPV cells. The calculations of the volume of the battery and area of the cells are described in Supplementary Information D. Figure 4(b) shows the cost per energy capacity (CPE) as a function of $T_{BB}$ and the stored thermal energy, with Si mass ranging from $10^3$ kg to $10^5$ kg, corresponding to a maximum energy capacity of ~0.11 GWh. The CPE is defined as:

$$\mathrm{CPE} = \frac{\mathrm{CapEx}}{Q_h} = \frac{\mathrm{CapEx}}{Q_{sen} + Q_{lat}} = \frac{\mathrm{CapEx}}{\int_{T_{ini}}^{T_h} m_{Si} \cdot c_p(T) dT + mL_{heat}}, \quad (1)$$

where $Q_h$ is the total energy stored in the thermal battery. The CapEx denotes the total capital expenditure, which comprises the primary material costs (Si and SiC mass) and an additional 20% factor [46] to account for balance-of-plant items such as insulation, labor, and other indirect costs. $Q_{sen}$ and $Q_{lat}$ are the sensible and latent heat components, $T_{ini}$ and $T_h$ are the initial and emission temperatures, $m_{Si}$ is the Si mass, $c_p(T)$ is the temperature-dependent specific heat of Si [47], and $L_{heat}$ is the latent heat of fusion. The volumes of Si and SiC, CapEx, and CPE calculation methods are detailed in Supplementary Information D. Lower CPE is crucial because it directly reduces the LCOE for TPV systems [48, 49]. As $T_{BB}$ and stored $Q_h$ increase, CPE significantly decreases due to the strong temperature dependence of sensible heat and the latent heat contribution near the Si melting point ($T_{melt} = \sim 1410\ °C$). This leads to a pronounced reduction in CPE, as the increase in stored energy outpaces system cost growth. Maximizing energy storage and minimizing CPE at high emission temperatures are highly advantageous for achieving high-power and high-efficiency TPV operation. Table 1 lists the parameters used for techno-economic analyses. Above the melting point, challenges such as chemical reactivity, containment, and thermal stress associated with liquid Si may increase system costs; however, detailed analysis of these effects is beyond the scope of this study. Note that a key economic characteristic of the TPV system is that the CPE exhibits diminishing returns as storage capacity increases. This behavior occurs because the fixed power-capital costs (e.g., TPV cells, in $/kW) are amortized over a larger energy capacity (kWh), while the energy-capital costs (e.g., the storage medium, in $/kWh) are scalable. Consequently, the CPE does not decrease indefinitely but asymptotically approaches a floor value determined by the irreducible energy-related component costs, as illustrated in the inset which plots the relationship between $m_{Si}$ and CPE.

The cost per power (CPP) is defined as:

$$\text{CPP} = \frac{\text{CapEx}}{\text{Total power}} = \frac{\text{Cost}_{fab} + \text{Cost}_{heatsink}}{P_{out,eff} \times \text{Cell Area} \times}, \qquad (2)$$

where $\text{Cost}_{fab}$ and $\text{Cost}_{heatsink}$ denote the fabrication cost and the cooling cost for the TPV

module, respectively. The calculations include various component costs, such as manufacturing of air-bridge TPV cells, maintenance, and heatsink. The TPV cell area is scaled with the thermal battery volume. The Cost$_{heatsink}$ is calculated based on a unit heatsink cost (50 $/kW) [50] divided by the total effective dissipated power, which is expressed as:

$$P_{\text{diss,eff}} = P_{\text{diss,static}} + P_{\text{diss,dynamic}}. \tag{3}$$

Here, $P_{\text{diss,eff}}$ accounts for not only static power loss but also the additional resistive heating ($I^2 R_s$) during discharge operations. The static component, $P_{\text{diss,static}}$, is the intrinsic cell loss, defined as $P_{\text{abs}} - P_{\text{out}}$ (where $P_{\text{abs}}$ is the total power absorbed by TPV cells) – calculated from $P_{\text{out}}$ shown in Fig. 3. The dynamic component, $P_{\text{diss,dynamic}}$ is calculated as $P_{\text{out}} \times f_{\text{loss}} \times (t_{\text{d,ref}}/t_{\text{d}})^2$ where $t_{\text{d,ref}}$ is the reference discharge time (10 hr). The $f_{\text{loss}}$ is the fraction of the Joule heating power loss to the input power, expressed by $I_{\text{mpp}}^2 \cdot R_s / (P_{\text{out}} + P_{\text{diss,static}})$ where $I_{\text{mpp}}$ is the current in the TPV cell at the maximum power point. Therefore, during the operation, the effective output power ($P_{\text{out,eff}}$) is reduced from the reference output power ($P_{\text{out}}$) by the amount of $P_{\text{diss,dynamic}}$.

Figure 4(c) and (d) show the CPP of the air-bridge InGaAs ($R_s$ = 7 mΩ·cm$^2$) and Si ($R_s$ = 80 mΩ·cm$^2$) TPV cells. The analysis was performed for various system scales, which are defined by $m_{\text{Si}}$ in the thermal battery (10$^3$ to 10$^5$ kg). The $m_{\text{Si}}$ directly determines the required TPV cell area and, consequently, the total $P_{\text{out,eff}}$ of the system. Assuming TPV systems of the same scale with $m_{\text{Si}}$ = 10$^5$ kg, the Si-based system generates up to ~0.8 MW, whereas the InGaAs-based system generates a higher power of ~8 MW. However, the higher power generation of the InGaAs cell also leads to greater dissipated power. This increased thermal load necessitates a more substantial and costly heatsink, which in turn increases the CPP of the InGaAs-based system. As a result, despite the large difference in power output, both the Si and InGaAs systems ultimately exhibit a similar CPP range. This illustrates a critical trade-off: although elevating $T_{BB}$ reduces the CPP by increasing power density, it also increases the dissipated power and thus the required heatsink expenditure. Therefore, maintaining a low heatsink cost is critical to fully realizing the economic benefits of high-temperature operation and achieving competitive CPP.

Figure 4(e) represents the CPP ($m_{Si}$ = $10^5$ kg and $T_{BB}$ = 1800 °C) and the corresponding $P_{out,eff}$ as a function of $t_d$. The change of $P_{out,eff}$ ($\Delta P_{out,eff}(t_d)/P_{out,eff}(t_d = 10\text{hrs})$) decreases sharply due to the significant increase in $P_{diss,dynamic}$. This reduction in $P_{out,eff}$ leads to a drastic increase in the CPP for the InGaAs system, despite its higher absolute $P_{out}$. This result indicates that designing the thermal battery for a sufficiently long discharge duration is critical for the economic viability of TPV technology based on III-V materials.

Fig. 4(f) and (g) compare the $P_{out,eff}$ and CPP of the InGaAs TPV cell with Si TPV cells having different $R_s$ (7, 30, 80 mΩ·cm²). The full colored graphs of CPP for each case as a function of $T_{BB}$ and $P_{out}$ are shown in Fig. S7. At a practical but high $R_s$ of 80 mΩ·cm², the Si TPV cell shows a relatively high CPP against the InGaAs cell due to its lower $P_{out,eff}$. However, as $R_s$ is reduced, the CPP of the Si cell improves, becoming comparable or lower to that of the InGaAs cell. For instance, reducing $R_s$ to 30 mΩ·cm² increases the maximum $P_{out,eff}$ to 2 MW ($m_{Si}$ = $10^5$ kg, $T_{BB}$ = 1800 °C). This result clearly indicates that lowering the $R_s$, and particularly the $R_c$ is the most critical challenge to making Si TPV economically competitive with its III-V counterparts in terms of CPP.

As illustrated in Fig. 1(a), the thermal battery is charged by both the solar PV and CSP. For cost comparison between Si and InGaAs-based TPV cells, the system is simplified by excluding items such as auxiliary boilers and additional thermal energy storage (TES) for backup. During operation, the system is self-sustaining without requiring external electricity.

The corresponding levelized cost of electricity storage (LCOS) is calculated as [48]:

$$\text{LCOS} = \frac{1}{N_{cy}} \left( \frac{CPP^+_{PV}}{\eta_{rt} t_c} + \frac{CPP^+_{CSP}}{\eta_{rt} t_c} + \frac{CPP^+_{in}}{\eta_{rt} t_c} + \frac{CPP^+_{out}}{t_d} + \frac{CPE^+}{\eta_{out}} \right), \qquad (4)$$

where $N_{cy}$ is the annual number of charging ($t_c$) and discharging ($t_d$) cycles. The cost terms of $CPP^+_{PV}$, $CPP^+_{CSP}$, $CPP^+_{in}$, $CPP^+_{out}$, and $CPE^+$ are annualized via the capital recovery factor (crf), $\frac{r}{1-(1-r)^{-n}}$ ($r$: the discount rate, $n$: the system lifetime). The round-trip efficiency is defined as $\eta_{rt} = \eta_{in} \times \eta_{out}$, where $\eta_{in}$ and $\eta_{out}$ are the input and output conversion efficiency, respectively. Each can account for heat losses, and $\eta_{in} = \eta_c / (1 + k_{loss} t_c/(t_c+t_d))$ and $\eta_{out} = \eta_d (1 - k_{loss} t_d/(t_c+t_d))$ where $k_{loss}$ denotes the fraction of the total energy capacity dissipated during each operational cycle as a result of heat leakage across the thermal insulation [48]. The $CPP^+$ for PV and CSP is calculated by:

$$\frac{\text{Unit cost (\$/m}^2)}{\text{Generated power density (W/m}^2)} \times \frac{\text{crf}}{\eta_{rt} \cdot t_c}, \qquad (5)$$

where the relevant parameters are in Table 1. The total required energy ($Q_{total}$) from the PV and CSP charging systems is calculated based on the assumption that the PV and CSP systems contribute equally to charging the thermal battery.

Figure 5(a)–(c) provide a detailed breakdown of the LCOS for the self-sustaining TPV system with air-bridge InGaAs cells (assuming $k_{loss} = 0.1$, $t_c$ and $t_d = 10$ hrs, and charging efficiency, $\eta_{ch}$, of 90 %). The analysis is presented in three scenarios: (a) including the full material costs of the Si storage medium and the SiC vessel, (b) excluding the SiC vessel cost, and (c) excluding all thermal battery material costs. The key finding is that the high LCOS of the TPV system is overwhelmingly dominated by the capital cost of the thermal battery. As shown in Fig. 5(c), when the material costs of Si and SiC are excluded, the remaining system LCOS

becomes competitive with other technologies like Li-ion, PHS, and hydrogen (LCOS ranges of Li-Ion batteries: $0.1 – $0.4/kWh [48, 51], pumped hydro storage (PHS): $0.05 – $0.2/kWh [51], and hydrogen energy storage: $0.1 – $0.3/kWh [52]). However, when the substantial material costs of Si or SiC are included (Fig. 5(a) and 5(b)), the LCOS of the TPV system is significantly higher than the other technologies. As illustrated in Fig. 5(d), operational inefficiencies such as heat loss further increase the LCOS, exacerbating the economic challenge. This collectively demonstrates that from a pure storage cost perspective, achieving economic competitiveness against established storage technologies is critically dependent on reducing the high capital expenditure required for the thermal battery installation.

Figure 6(a) presents the LCOE for the self-sustaining InGaAs TPV system as a function of $T_{BB}$ and the total energy generated with the system lifetime of 25 yrs [53]. The LCOE is calculated by dividing the sum of the lifetime CapEx and operational (OpEx) expenditures by the total energy generated over the system lifetime (accounting for practical factors such as an annual TPV cell performance degradation, $d_r$, of 5% [48, 54]). The details of the calculations are described in Supplementary Information G. Figure 6(b) represents that the LCOE of the InGaAs-based TPV system exhibits a critical dependence on the heat loss rate of the system. While the LCOE remains relatively low for heat loss rates up to approximately 50–70%, it increases sharply beyond this threshold. This behavior is a direct consequence of the LCOE definition (Total Cost / Total Energy); as heat loss increases, the total net energy generated decreases, causing the LCOE to increase hyperbolically. The loss of stored energy is independent of cell degradation, meaning that the suppression of heat loss is important for competitive LCOE of the TPV system. Figure 6(c) further explores the LCOE over various system lifetimes with three different heat loss rates (10 %, 20 %, and 50 %) ($d_r$ was fixed to 5%). A key finding is that the LCOE shows diminishing returns with lifetime, beginning to saturate after 15–20 years. The combined analysis in Fig. 6(b) and (c) reveals a crucial insight: while a sufficiently long lifetime is necessary to amortize the initial capital expenditure, for systems designed for long-term operation (>20 years), the heat loss rate becomes the dominant factor in determining the final, saturated LCOE value. A lower heat loss rate establishes a fundamentally lower cost floor, implying that further economic improvements are more effectively realized by enhancing the thermal insulation of the system rather than by simply extending its operational lifetime.

Figure 6(d) provides further investigations of the LCOE with a massive-scale system ($m_{Si}$ varies from $10^5$ kg to $10^7$ kg) indicates that even a Si-based TPV system with a practical $R_s$ (e.g., 80 mΩ·cm$^2$) can achieve an LCOE in the range of \$0.29/kWh. As stated above, reducing the CPE, the LCOE can be less than the prediction. Eventually, the optimized TPV systems show an LCOE comparable to other technologies such as PV (0.03–0.08 \$/kWh) [55], CSP (0.1–0.15 \$/kWh) [56], Li-ion (0.12–0.19 \$/kWh) [57], and gas-turbine (0.15–0.32 \$/kWh) [58, 59] (see Fig. S8 in Supplementary Information). Note that $P_{diss}$, defined in Equation (3), affects the LCOE. This economic competitiveness is particularly pronounced for large-scale power grid applications, especially at the gigawatt-hour (GWh) scale and beyond, where long-duration, dispatchable power is most critical. Thus, the primary economic advantage of the TPV system lies in its function as a self-sustaining power generation asset. By eliminating the need for external fuel or electricity, it offers a unique pathway to providing zero emission, on-demand power, a role currently filled by fossil fuel generators. In addition, while the LCOE of an optimized InGaAs cell may remain lower than that of a practical Si cell, proven, immense manufacturing scalability presents a lower-risk and highly promising pathway for deploying this technology at the grid scale required for a zero-emission future. Figure S9 provides additional estimations of the LCOE for scenarios with a reduced $P_{diss}$ and different $t_c$ and $t_d$ times. Note that while $P_{\text{diss}}$ is a realistic constraint, a hypothetical reduction in this parasitic loss would enhance the round-trip efficiency, leading to a remarkable reduction in the LCOE. For example, as shown in Fig. S9(c), the LCOE for the Si-based TPV cell with an $R_s$ of 80 mΩ·cm² can be reduced to less than $0.2/kWh.

**Discussion and Conclusion**

This study establishes the techno-economic viability of a novel self-sustainable TPV architecture that functions as a standalone power generation asset by integrating solar charging, thereby eliminating the reliance on external electricity. The viability of this architecture was evaluated for both established III-V and low-cost Si TPV cells. We identified that for materials like silicon to become competitive, critical design principles – a thinned wafer (~50 μm) and a low series resistance (<30 mΩ·cm²) – are required to unlock conversion efficiencies exceeding 35%. While system-level analysis indicates an ideal InGaAs-based system may exhibit a marginally lower LCOE, this does not fully capture the practical risks associated with manufacturing scalability. A massive-scale system analysis reveals that even a Si-based TPV with a practical $R_s$ can achieve an LCOE competitive with conventional gas-turbine power plants. Ultimately, this work demonstrates that the self-sustainable TPV architecture provides a compelling economic pathway to serve as a source of zero-emission, dispatchable power. The analysis highlights that leveraging the proven, immense manufacturing scalability associated with silicon presents a lower-risk and highly promising route for deploying this technology at the grid scale required for a zero-emission future.

**Materials and Methods**

**FTIR measurements and Drude model** A thin Ti/Au (10nm/100nm) layer was deposited on a Si wafer using e-beam evaporation. Subsequently, the prepared ground Si wafers with different thicknesses were attached to the Au layer using silver paste to form air bridges. To examine the dependence of parasitic photon losses on Si wafer thickness, FTIR measurements (ThermoFisher Scientific, Nicolet RaptIR+ FTIR Microscope) were performed across wavelengths ranging from 1.3 μm to 15.4 μm. The measured data were used to calibrate the Drude model using the equation of extinction coefficient, $k_r = \frac{1}{4\pi}(\lambda^\gamma C n)$, where $\lambda$ is the wavelength, $\gamma$ and $C$ are fitting parameters dependent on the sample absorption, and $n$ is the free carrier concentration [29, 60]. This enabled a quantitative assessment of FCA-induced photon loss in the Si wafers.

**TCAD simulation** A TCAD suite (Sentaurus, Synopsys) is employed to generate the Si TPV model structure and simulate J–V characteristics. The simulations incorporate Shockley–Read–Hall recombination, surface recombination, field- and doping-dependent carrier mobility, and are coupled with the thermodynamic model to solve for the lattice temperature. This approach accounts for self-heating effects, where Joule heating generated in the device is dissipated through a thermal contact (Thermode) defined at the cell bottom. A thermal boundary condition with a surface resistance of 1.5 cm$^2$K/W, consistent with the diode model, is applied to this thermode. The transfer matrix method is coupled to the simulation to calculate the photocurrent under blackbody radiation conditions, where the blackbody spectrum is manually input based on the Stefan–Boltzmann law. The device geometry, contact definitions, doping profiles, and material parameters are specified as described in Supplementary Information B.

**Si grinding process** Single-side polished and p-type doped (Boron 10$^{16}$ /cm$^3$) Si wafers with an initial thickness of 250 μm were thinned to target thicknesses of 200 μm, 100 μm, and 50 μm using a back grinding process. The wafers were mounted with the polished side facing down, and mechanical grinding was performed from the back side using a series of progressively finer abrasive wheels to achieve the desired thicknesses. After grinding, the wafer thickness was confirmed by micrometer measurement. No additional chemical-mechanical polishing (CMP) or etching steps were applied following back grinding. Figure S10 presents the resultant variations in silicon wafer thickness observed by an optical microscope (Olympus BX51).

**Series resistance measurement** The dark $J$-$V$ measurement in Supplementary Information Fig. S3(d) was performed on a conventional Si photodiode (P&L SEMI CO., LTD., South Korea) with a metal (Al 30nm/Ti 10nm/Pt 30nm/Au 3000nm) contact area of 0.09 cm$^2$. The measurement was performed using a Keysight B1505A Power Device Analyzer equipped with a B1512A high-current source-measurement unit. The measurement range was limited by the module's maximum specifications of 20 V and 20 A (pulsed), and its overall maximum power output.

**Acknowledgement**

We acknowledge the valuable support of Dr. Kang Bok Ko and Professor Cheol-Jong Choi (School of Semiconductor and Chemical Engineering, Jeonbuk National University) for their assistance with the Si grinding process. The fabrication of all devices was performed at the cleanroom facility of the Semiconductor Physics Research Center, Jeonbuk National University. This research was supported by the Global-Learning & Academic Research Institution for Master's, Ph.D. students, and Postdocs (LAMP) Program of the National Research Foundation of Korea (NRF), funded by the Ministry of Education (No. RS-2024-00443714). This work was also supported by the Korea Basic Science Institute (National Research Facilities and

Equipment Center) grant funded by the Ministry of Education (grant No. RS-2024-00436672).

## Author contributions

J. Lim conceptualized the idea and method, supervised the project, performed simulations, wrote the initial draft, and revised the manuscript. J. Lim and S. Lee designed the experiments and analyzed the data.

## Competing interests

The authors do not have any conflict of interest to disclose.

**Figures and Tables**

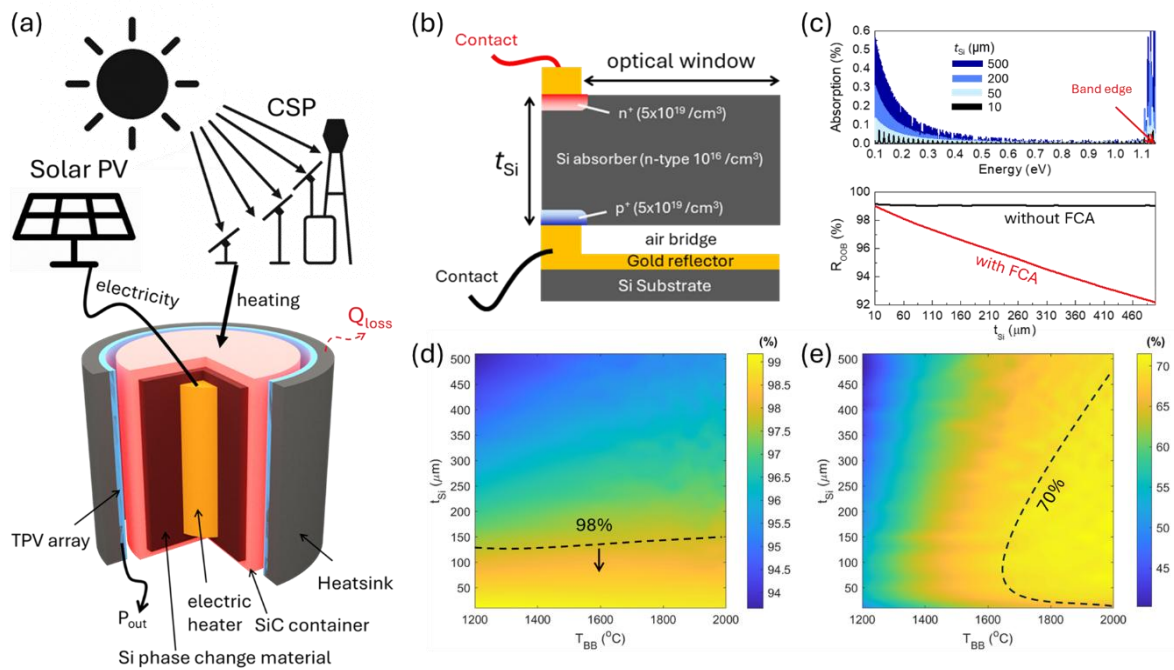

**Figure 1.** (a) Schematic illustration of the integrated thermophotovoltaic (TPV) system, which combines a solar photovoltaic (PV) module and a concentrated solar power (CSP) unit for charging the thermal battery, with electricity generation via a TPV cell array. This architecture represents both a sustainable and renewable energy system. (b) Schematic of the unit structure of the Si air-bridge TPV cell, where

$t_{Si}$ denotes the thickness of the Si absorber. (c) Fourier-transform infrared (FTIR) measurements (top panel) and simulations (bottom panel) on Si absorber/air/Au/Si samples with varying $t_{Si}$. (d) Simulated out-of-band reflectance ($R_{OOB}$) as a function of blackbody temperature ($T_{BB}$) and $t_{Si}$. (e) Simulated spectral efficiency (ature$SE$) as a function of $T_{BB}$ and $t_{Si}$. Dashed lines indicate regions where $R_{OOB}$ exceeds 98% and $SE$ exceeds 70%, demonstrating performance comparable to III-V thin-film air-bridge TPV cells.

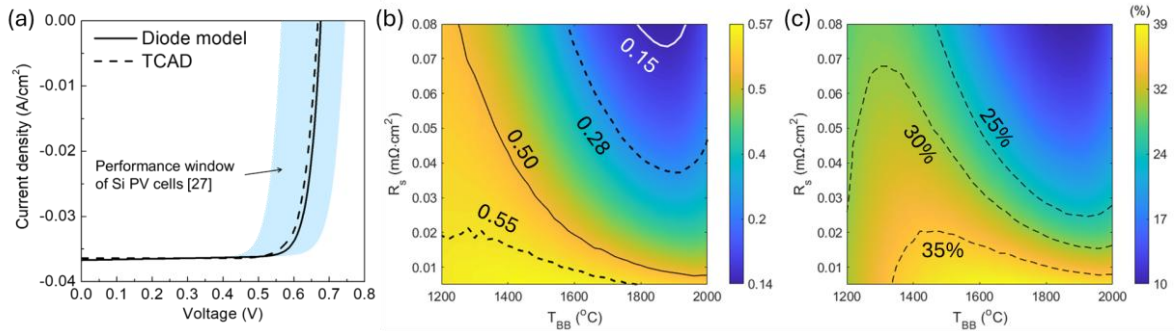

**Figure 2.** (a) Simulated current-density versus voltage ($J$-$V$) characteristics of Si TPV cells using TCAD (Sentaurus, Synopsis) and a diode model [26]. The performance window indicates the variation of open-circuit voltages ($V_{oc}$) owing to different radiative recombination processes. (b) Contour plot for carrier management – Voltage factor ($V_F$) multiplied by fill factor ($FF$) – as a function of series resistance ($R_s$) of the Si TPV cell and $T_{BB}$. The $t_{Si}$ was fixed to 50 μm. (d) Conversion efficiency (%) as a function of $R_s$ and $T_{BB}$, calculated for $t_{Si}$ = 50 μm.

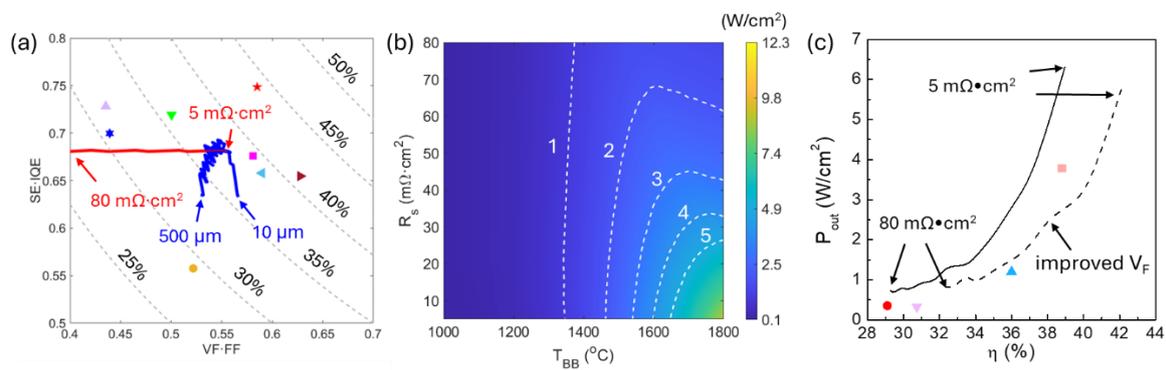

**Figure 3.** (a) Figure of Merit (FoM) map for TPV performance, showing power conversion efficiency contours as a function of $V_F \cdot FF$ (voltage factor × fill factor) and $SE \cdot IQE$ (spectral efficiency × internal quantum efficiency). Dashed lines indicate constant conversion efficiency levels from 25% to 50%. The red trajectory illustrates the effect of $R_s$ reduction from 80 mΩ·cm² to 5 mΩ·cm². The blue curve represents the effect of varying Si absorber thickness from 10 μm to 500 μm. Yellow symbols indicate published TPV cell performance data from the literature. (b) Contour plot of simulated $P_{out}$ of the Si TPV cell with $t_{Si}$ = 50 μm as a function of $T_{BB}$ and $R_S$. The dashed contour lines indicate constant $P_{out}$, showing the strong dependence of power generation capability on both $T_{BB}$ and $R_s$. (c) Figure of merit of InGaAs-based TPV cells (symbols) [8, 14, 27, 45]. The solid line represents the performance of the Si TPV cell with $R_s$ varying from 5 mΩ·cm² to 80 mΩ·cm². The dashed line shows the projected improvement with enhanced $V_F$ [26].

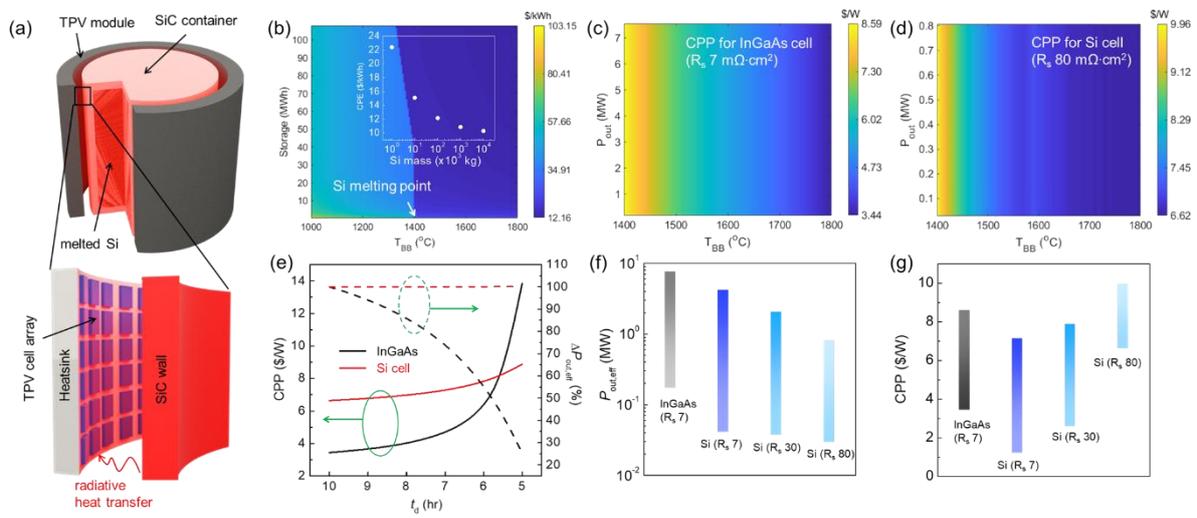

**Figure 4.** (a) Schematic diagram of the thermal battery integrated with a TPV cell array, illustrating radiative heat transfer through the SiC wall. (b) Contour plot of the cost per unit energy (CPE, $/kWh) as a function of storage energy and emitter temperature, $T_{BB}$, with the phase transition point of Si indicated. Contour maps of the cost per power (CPP, $/W) as a function of output power density ($P_{out}$) and $T_{BB}$ for (c) an InGaAs TPV system with $R_s$ = 7 mΩ·cm² and (d) a Si TPV system with $R_s$ = 80 mΩ·cm² as a conventional value. (e) CPP for the InGaAs and Si cells as a function of discharging time ($t_d$). (f) and (g) Comparison of effective output power ($P_{out,eff}$) and CPP for the InGaAs and Si TPV cells with various $R_s$.

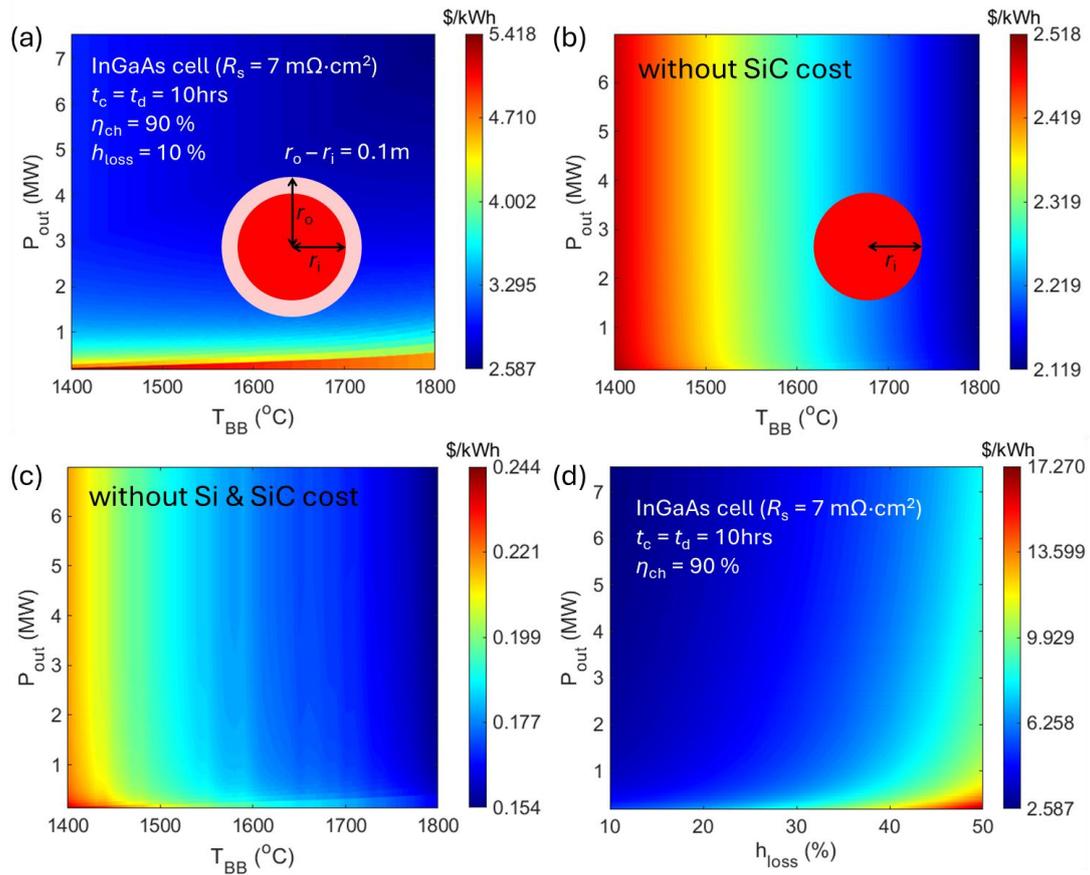

**Figure 5.** The calculated levelized cost of storage (LCOS) considering (a) full material costs for the Si storage medium and SiC vessel with the different radius (0.1 meter between the outer radius for SiC, $r_o$, and the inner radius for silicon, $r_i$), (b) excluding the SiC vessel cost, and (c) excluding all thermal battery material costs. (d) The impact of heat loss (0% to 50%) on the LCOS for the full-cost scenario. The details of the techno-economic assumptions are described in Table 1 and Supplementary Information F.

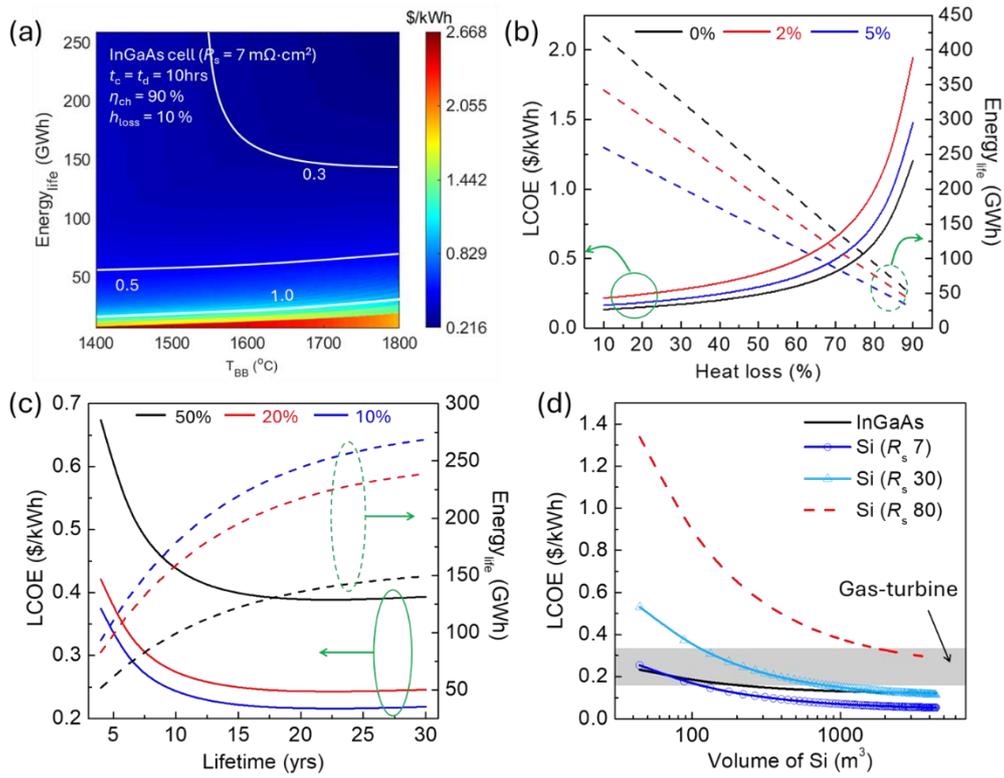

**Figure 6.** Levelized cost of electricity (LCOE) analysis of the self-sustaining TPV system. (a) LCOE for the InGaAs-based TPV system as a function of $T_{BB}$ and the total energy generated, scaled by the Si mass in the thermal battery ($10^3$ to $10^5$ kg). (b) The impact of heat loss on the LCOE with different annual TPV performance degradation rates (0%, 2%, and 5%). (c) The LCOE as a function of the system's operational lifetime with different heat loss rates of (10%, 20%, and 50%). (d) LCOE for the InGaAs and Si-based TPV systems with different $R_s$ (unit: m$\Omega \cdot$cm$^2$) as a function of the volume of Si in the thermal battery. The calculations consider the mass of Si from $10^5$ kg to $10^7$ kg. The detail of the calculations are described in Supplementary Information D (thermal battery) and G (LCOE calculations).

**Table 1.** Parameters and abbreviations used in techno-economic assumptions such as cost per energy capacity, cost per power, levelized cost of energy, and levelized cost of electricity storage.

| Parameter (unit) | Value |
|---|---|
| Si mass ($m_{Si}$) (kg) | $10^3$ to $10^5$ <br> *$10^5$ to $10^7$ in Fig. 6(d) |
| Si melting point, $T_{melt}$ (°C) | 1410 |
| Latent heat of Si, $L_{heat}$ (kJ/kg) | 2000 |
| Cost of Si ($/m³) | 20000 |
| Cost of SiC ($/m³) | 35000 |
| Si gravimetric density (kg/m³) | 2263 |
| SiC gravimetric density (kg/m3) | 3210 |
| Cost$_{heatsink}$ ($/kW) | 50 |
| $t_c$ (hr) | 10 |
| $t_d$ (hr) | 10 |
| Lifetime, n (yr) | 25 |
| number of cycles per year, $N_{cycl}$ | 730 |
| Annual device degradation rate, $d_r$ (%) | 5 |
| charging efficiency, $\eta_{ch}$ (%) | 90 |
| energy input efficiency, $\eta_{in}$ (%) | calculated as a function of $\eta_{ch}$, $h_{loss}$, $t_c$, and $t_d$ |
| energy output efficiency, $\eta_{out}$ (%) | calculated as a function of $\eta_{cell}$, $h_{loss}$, $t_c$, and $t_d$ |
| operating expenditure, OpEx ($/kW-year) | 12.5 (provided in Table S4) |